\documentclass[twocolumn,aps,superscriptaddress]{revtex4-1}

\usepackage{graphicx}
\usepackage{latexsym}
\usepackage{amssymb}
\usepackage{amsmath}
\usepackage[english]{babel}

\begin{document}

\title{Observational signatures of the giant planets collisions}

\author{V. I. Dokuchaev}\thanks{e-mail: dokuchaev@inr.ac.ru}
\author{Yu. N. Eroshenko}\thanks{e-mail: eroshenko@inr.ac.ru}
\affiliation{Institute for Nuclear Research of the Russian Academy of
Sciences \\ 60th October Anniversary Prospect 7a, 117312 Moscow, Russia }

\date{\today}

\begin{abstract}
We consider observational signatures of the collisions and partial destructions of giant exoplanets at the chaotic stage of the planetary systems evolution. The rate of these collisions in the Galaxy is estimated to be $\sim0.01-1$~yr$^{-1}$. In the inelastic collision with a small relative velocity and small impact parameter, the planets may sufficiently lose their kinetic energy and merge together. Otherwise, if the planet is experienced a shallow tangential collision, it flews safety away. At the same time, the planets loss some part of their gas envelopes during mutual collisions. Collisions of the giant planets must be accompanied also by the radiation of energy from the radio to optical bands. The optical and near UV flashes result from the collision heating of the planet surface layers. Additionally, the collision compression and collapse of the powerful magnetosphere of giant planets are possible, with a successive generation of the radio bursts. According to our estimations, the corresponding rate the radio bursts is $\sim0.01-1$ bursts per year with the maximum spectral flux $\sim30 $~mJy at the frequencies $\sim3 $~GHz and with a duration of the each burst $\sim1.5$~hours. These signals are available for registration by the existing radio telescopes. 
\end{abstract}

\maketitle


\section{Introduction}

Recently a lot of exoplanets --- the planets outside the solar system in the orbits around the distant stars, have been found. Most of the discovered planets belong to the class of giant planets, like Jupiter, because of the observational biases and limitations. A few of the solid-surface extrasolar planets of the Earth type are also been detected. Now, it is already evident that the planetary systems are not rare, but instead of, are the typical astronomical phenomena. About 8-23\% of stars contain Jupiter-like planets (0.3-10 of Jupiter mass) with an orbital semi-axis $a=0.5-10$~AU \cite{Cas12}. Some stars contain multiple gas giants, e.\,g., the system HR~8799 has four of these planets \cite{Mar10}.

The planets in the Solar and the other planetary systems have being born in the gas-dust disks, resulting from the compression of the protoplanetary clouds along the axes of common rotation. The terrestrial planets formed by the aggregation of dust particles and the larger planetesimals. The giant planets are formed probably by the accretion of gas onto the solid seeds with a mass of the order of a few Earth masses or in the alternative process, from the gas condensation due to the gravitational instability in the protoplanet disk \cite{Moretal10}. After the starting of nuclear reactions in the central star, the light elements remaining in the gaseous phase were carried out to the periphery by the stellar wind, with the exception of material which is already condensed into the protoplanets.

The protoplanets can scatter gravitationally, collide, merge or destruct during the various stages of the planetary system evolution \cite{ForChi07,RayBarMan08,Chaetal08,LiAgnLin10}. The catastrophic collisions are possible especially at the early stage of planet formation, when mutual scattering of planets are most often, and the protoplanet orbits are rather chaotic and unstable. The process of planetesimal collisions and merges is considered as the mechanism of aggregation and growth of the terrestrial planets and as the central stony or metal cores of giant planets. The collisions with a successive merging or destruction of the terrestrial planets were considered, for example, in \cite{GenKokIda11,SteLei11-1,SteLei11-2}. Collisions and merges of the giant planets may also be the important factors in the early evolution of at least in some of protoplanetary systems \cite{LiAgnLin10}. There are several mechanisms for the appearance of the overlapping planetary orbits, resulting in the catastrophic collisions. These are the collisions at the early chaotic stage, the convergence and intersection of orbits in their slow evolution (migration) and the resonant and three-body interactions.

The consequences of the mutual collisions of giant planets will depend, in particular, on their internal structure, the ratio of planet masses, the relative velocities, the impact parameters and the angular momentums. The corresponding consequences of collisions for terrestrial planets with the solid surfaces were considered in \cite{GenKokIda11}. For the giant planets, the depth of interpenetration of the gaseous (or liquid) shells of two planets is very important, because the density in the outer layers increases inward very  sharply. In the inelastic collision with a small relative velocity, if the collision inveigles a large fraction of mass, the planets lose a sufficient part of there relative kinetic energy to merge into a single object. Otherwise, the planets will experience only a shallow tangential collision and flew safety away. In any case, some amount of the planet material may be released, especially in the region in between the planets, where the mutual gravity influence is compensated. A gaseous component from the external layers will quickly disperse after collision. The intriguing problem is a fate of the intermediate regions of giant planets, which are probably composed of the liquid metallic hydrogen \cite{Ash68,Zhar78,Hab87}. The disturbance of the metallic hydrogen region may drastically change the magnetic dynamo operation, resulting in, probably, a very powerful electromagnetic energy release.

Collisions of giant planets must be accompanied by the energy release from the radio to optical bands. The optical and near UV radiation are the results of the collision heating, or, in other words, the transformation of mechanical energy into the heat. The compression and destruction of the planet magnetospheres lead to the generation of the powerful bursts of the radio waves at the cyclotron frequencies. The corresponding signals are available for registration. The magnetospheric emission from the extrasolar planets and perspectives of their detection were discussed in the literature, see e.\,g. \cite{ReiChr10,Lazetal10} and references therein. This emission (which is observable for the case of the Solar system planets) is a result of the interaction between hosting stellar wind and the planets magnetospheres. The anticipated signal from the Jupiter-like planet has the frequency $\sim40$~MHz and with a power $\sim10^{18}$~erg~s$^{-1}$. In this paper we predict a harder and more energetic radio bursts resulting from the giant planets collisions.


\section{Rate of giant planet collisions}
\label{ratesec}

At the initial stage of the planetary system formation, the planets have a rather chaotic orbits \cite{RayBarMan08} and they are often intersecting. A a result, the planets are scattered, collided, partially destroyed and merged. The possible explanation of specific configuration of the solar system may be in the early existence of the fifth giant planet, which was ejected from the solar system in the scattering event \cite{Nes11}. The important process for the formation of planetary systems is the so-called  effect of ``orbit migration'' --- the slow orbit change due to interaction (exchange of the angular momentum) with the gaseous disk (see e.\,g. \cite{RayBarMan08,AylBat10}). For orbits with a different ellipticity and different inclination to the ecliptic, the orbit migration can lead to the intersections and collisions of planets.

The collision cross-section of two spherical massive bodies is \cite{LL-1}:
\begin{equation}
\sigma_{\rm coll}=\pi(r_1+r_2)^2\left[1+\frac{2G (m_1+m_2)}{(r_1+r_2)v_{\rm rel}^2}\right],
\label{sigma}
\end{equation}
where $r_{1,2}$ and $m_{1,2}$ are the planet radii and masses, and $v_{\rm rel}$ is their relative velocity. The first term in the brackets corresponds to a purely geometrical collision cross-section, while the second term takes into account the gravitational focusing. If we consider the process of the destruction of the inner layers of a gas-giant planets, then radii $r_{1,2}$ in (\ref{sigma}) should be considered as the radii of these layers, while masses $m_{1,2}$ remain equal to the total masses of the planets. For estimation, we assume that a protoplanetary disk contains $N$ giant planets inside a characteristic heliocentric radius $L$ and thickness $2H$. The thickness of the protoplanetary disk can be estimated from the relation $H\simeq\alpha L$, where $\alpha$ is the characteristic angle of the orbit inclination to the median plane.

Let us consider at first the case of chaotic orbits. We do not take into account the dependence of all involved quantities on the distance from the star. We also consider only the average values of all characteristic parameters. The characteristic  planet collision time-scale, having the number density $n_2\sim N/(\pi L^3\alpha)$, is
\begin{equation}
t_{\rm coll}^{(1)}\sim\frac{1}{n_2\sigma_{\rm coll}v_{\rm rel}}\sim2.4\times10^5\mbox{~yr},
\label{tcol}
\end{equation}
where numerics correspond to the following choice of parameters: $\alpha=2.5^\circ$, $L=9$~AU, $v_{\rm rel}\simeq 13.1$~km~s$^{-1}$, $N=5$, $R=R_J\approx7\times10^4$~km, $M=M_J\approx10^{-3}M_\odot$.

The motion of protoplanets is gradually in time becoming a more regular one. In result, the convergence of the Keplerian orbits at the later stage can not be considered as an accidental. Consider now the case, when the protoplanet orbits are almost stabilized, with a remained orbit migration time-scale $t_{\rm mig}\sim10^5$~yr \cite{Waletal12}. The characteristic velocity of orbit migration is $v_{\rm mig}\sim L/(2t_{\rm mig})\simeq2\times10^{-4}$~km~s$^{-1}$. We introduce now the parameter of chaos:
\begin{equation}
\eta=\frac{v_{\rm mig}T_{\rm orb}}{r_p}\sim0.1\left(\frac{t_{\rm mig}}{10^5\mbox{~yr}}\right)^{-1},
\label{eta}
\end{equation}
where $T_{\rm orb}$ is an average orbital period of the planets, and $r_p=(\sigma_{\rm coll}/\pi)^{1/2}$ is the impact collision parameter. The randomness condition, $\eta>1$, implies that possible successive collisions are all independent of each other, and described by the  cross-section (\ref{sigma}). The collision time in this case is given by (\ref{tcol}). Otherwise, if $\eta<1$, the planet at the each next period will pass partially through the region of space where it was previously, and therefore the collision probability is suppressed. In other words, if $\eta<1$, the collision cross-section sweeps the volume of space $\sim tv_{\rm mig} 2r_p2\pi r_{\rm orb}$ during the time $t$, where $r_{\rm orb}\sim L/2$ is the average radius of the orbit. Using, as previously, the characteristic parameters of the protoplanetary disk, we estimate the collision time for the case of the orbit migration
\begin{equation}
t_{\rm coll}^{(2)}\sim \frac{1}{4\pi r_{\rm orb}r_p v_{\rm mig}n_2}\sim1.4\times10^6\mbox {~yr}.
\label{tcol2}
\end{equation}
Since the quantity (\ref{tcol2}) is of the order of magnitude greater than the corresponding characteristic time-scale of the migration, the probability of the planet collisions at the stage of the orbit migration would be $\sim1/10$. After the finishing of chaotic phase and stabilization of the orbits, the collisions are possible due to the slow changes of orbits under the influence of weak perturbations.

On the basis of smallness of the characteristic times-scales in the estimations (\ref{tcol}) and (\ref{tcol2}), it is reasonable to assume that $f_s\sim0.001-0.1$ for collisions of giant planets such as Jupiter per each star during the period of planet formation in the protoplanetary disk. Really, the simulations of \cite{MarWei02} show that the collision probability is of the order of 1-5\% due to the multiple (3 or more) planets interactions. If we took into account the conservative estimation that $\sim 10$\% of the star systems in the Galaxy contain Jupiter-like planets \cite{Cas12}, we obtain the lowest rate $f_s\sim0.001$, but in the optimistic scenario it's higher by about two orders. In the Galaxy, during the time $\sim10^{10}$~yr, there were $\sim10^{11}f_s$ collisions of protoplanets. In this estimation it is taken into account that stars in the Galactic disk are formed with a nearly steady rate during the history of the Galaxy \cite{BinTre94}. As a result, the rate of giant planet collisions in the Galaxy is $\sim 10f_s$~yr$^{-1}$. A similar estimation follows from the currently measured rate of star formation in the Galaxy, $3-4M_\odot$~yr$^{-1}$, if the planet collisions take place mainly during the early stage of planetary systems evolution.

\section{Merging or flying away of planets in collisions}
\label{mergefly}

There are two possibilities for the fate of planets after collision: (i) the merging into a single object or (ii) the grazing touching with some matter loss and a final flying away. For the total planet destruction collision, the relative velocity (which is of the order of the orbital velocity $v_{\rm orb}$) must exceed the escape velocity $v_2$ on the surface of the planet (for a solid planets the requested excess must be 2-3 times larger). For the Jupiter, orbiting a Sun-like star, the corresponding orbital radius for planet destruction is $r=GM_{\odot}/v_2^2=$0.25~AU. The condition $v_2<v_{\rm orb}$ is valid for the major part of known exoplanets --- ``hot Jupiters''. In the opposite case, $v_2>v_{\rm orb}$, the planets in the head-on collision will merge, and their fragments will combine finally into the single large planet.

The interesting are the tangential planet collisions, when planets are flying away after mutual close encounter, by experiencing the contact or grazing interactions only of their outer layers. Let us consider the case $v_2>v_{\rm orb}$. In the regions of geometric intersection of planets during collision, all collided matter will stop, and its kinetic energy is partially converted into the heat. The collided planets will flying away, and the stopped matter is partially gravitationally attracted  by the planets.

\begin{figure}[t]
\begin{center}
\includegraphics[angle=0,width=0.4\textwidth]{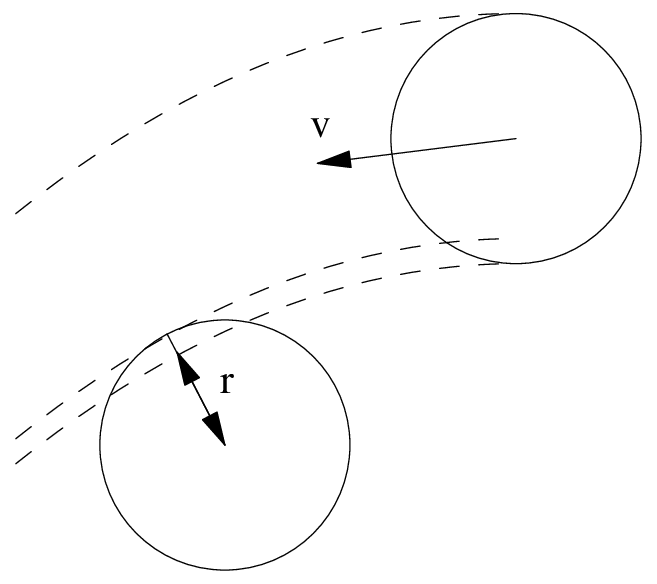}
\end{center}
\caption{The geometric configuration of the planets collision. The surface of the first planets moves relative to the center of the second one at the minimum penetration distance $r$. The ``hats'' on the collided planets enclosed by the dashed lines are grazed during the close encounter and partially ejected. } \label{conffig}
\end{figure}

Now we find a condition for the release of metallic hydrogen from the interior of the planets, during planet collisions. It is interesting, in view of the possible powerful electromagnetic energy release, during collision of this type. We also consider the possible metallic-hydrogen release in free space. This effect will provide in principle the unique information on the properties of the supposed metallic-hydrogen, which are unattainable in the modern laboratories. 

\begin{figure}[t]
\begin{center}
\includegraphics[angle=0,width=0.49\textwidth]{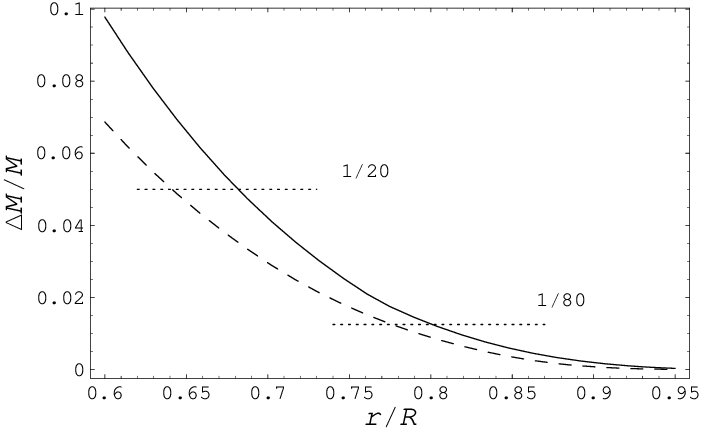}
\end{center}
\caption{The relative mass loss $\Delta M/M $ in dependence of the penetration radius $r$ (in units of the planetary radius $R$) of the giant planet with internal structure similar to the Jupiter. The horizontal dotted line indicates the $\Delta M/M\sim1/80$, when grazing collision attains the region of the metallic hydrogen. The second dotted line corresponds to the $\Delta M/M\sim1/20$, when collided planets are merged due to the loss of the relative kinetic energy. In the intermediate region the metallic hydrogen can be ejected. For comparison, the dashed curve shows the corresponding mass loss for the Saturn.} \label{mlossfig}
\end{figure}

To estimate the mass of the captured matter during the contact planet collision we approximate the collision area by the plane cut of the sphere, whose lower flat area is close to the center of the planet at a minimum penetration distance $r$, see Fig.~\ref{conffig}. The corresponding captured mass $\Delta M$, presented in the Fig.~\ref{mlossfig}, is
\begin{equation}
\Delta M=\pi R^3\int\limits_{r/R}^1dx(1-x^2)\rho(Rx),
\label{dm}
\end{equation}
where $\rho(r)$ is the matter density, depending on the distance from planet center $r=xR$, and $R$ is the planet radius. For the density $\rho(r)$ we take model calculations of the internal structure of Jupiter \cite{Zhar78}, shown in Fig.~35. The Fig.~\ref{mlossfig} shows the affected mass $\Delta M/M$ in dependence on the minimum distance from the planet center. The internal metallic hydrogen layers of the Jupiter will be affected by the collision, if $r/R<0.8$. The corresponding minimum release of the matter (the metallic hydrogen) in this case is
\begin{equation}
\Delta M/M\sim1/80.
\label{dm180}
\end{equation}

The relative velocity $v$ of the planets at the moment of the collision exceeds $v_{\rm rel}$ due to the gravitational focusing:
\begin{equation}
v=v_{\rm rel}\left[1+
\frac{2G (m_1+m_2)}{(r_1+r_2)v_{\rm rel}^2}
\right]^{1/2}
\simeq60\mbox{~km}\mbox{~s}^{-1}.
\label{vver}
\end{equation}
Let us assume for estimation that about half of the grazed mass is captured back by the gravitation of the planets. The corresponding energy loss for the attraction of this mass is $U\sim G(\Delta M/2)M/R$. This energy is equal to the work of pressure in the hydrostatically supported planet \cite{Hab87}. The planets do not merge into a single body, if $U$ is less than the initial kinetic energy of relative motion $\mu v_{\rm rel}^2/2 $, where $\mu=M_1M_2/(M_1+M_2)=M/2$ is the reduced mass and $v_{\rm rel}\sim v_{\rm orb}$. For collision of the Jupiter-like planets at the orbit of the radius $r_{\rm orb}$ around the solar-mass star, this criterion gives
\begin{equation}
\Delta M/M_\odot<\frac{R}{2r_{\rm orb}}\simeq4\times10^{-5}\left(\frac{R}{R_J}\right)\left(
\frac{r_{\rm orb}}{5.2\mbox{~AU}}\right)^{-1},
\label{dmku}
\end{equation}
i.~e., $\Delta M/M_J \sim1/20$.
Thus, there is a mass interval from (\ref{dm180}) to (\ref{dmku}), when matter ejection is possible. We provide here only the numerical examples, which is model dependent. This section contains only the rough order of magnitude estimations, which could be improved by the detailed numerical simulations.

The probability of collisions involving the grazing of the metallic hydrogen can be estimated from the cross-section (\ref{sigma}). This  cross-section is proportional to the first power of $(r_1+r_2)$ for the considered velocity conditions. From (\ref{dmku}) we find (see also Fig.~\ref{mlossfig}) that planets are merging with the probability 68\%. The release of the metallic hydrogen will occur with the probability $80-68=12$\%. The remaining probability 20\% corresponds to the tangent collisions, when planets are touched by only their outer atmospheres.

\section{Energy releases and radio bursts}
\label{radiosec}

Here we provide an order of magnitude estimation of the energy release at the collisions. The ejected metallic hydrogen is likely to have a very dispersed form, composed of the fragments and droplets of various sizes, which are expanded and evaporated very fast. The distribution of the fragments by sizes for the destruction of gravity-dominated bodies was calculated in \cite{SteLei11-1} and with a good accuracy has the power-law form $n(D)dD\propto D^{-(\beta+1)}dD$, with the index $\beta \sim3.5-4$. The upper metallic hydrogen layers in Jupiter have the temperature $T\sim 10^4$~K. Immediately after collision, the ejected fragments will experience the heating due to transfer of the kinetic energy the into heat. Indeed, the kinetic energy of matter is $E_0\sim\Delta Mv^2/2$. If the ejected material would be heated uniformly, then its temperature would rise by the amount $\Delta T\sim E_0/(\Delta MC)=v^2/(2C)\sim5\times10^4$~K, where $C$ is the specific heat capacity of metallic hydrogen. For $C$, we choose the heat capacity of liquid hydrogen from \cite{ChemE}
\begin{equation}
C(T)=6.86+0.66\times10^{-4}T+0.279\times10^{-6}T^2\mbox{~~kJ~kg$^{-1}$K$^{-1}$}
\label{teplc}
\end{equation}
with $T\sim10^4$~K. In addition, without the pressure support, the metallic hydrogen transforms into the dielectric phase with a possible energy release 290~MJ~kg$^{-1}$ \cite{PhysE}. Therefore, we expect the heating and the flashes of light up to the UV band. On the other hand, the expanded hydrogen droplets will cool after their ejection into the zero pressure environment. As a result, we expect the formation of very dispersed and partially ionized hydrogen clouds around the collided planets. This clouds will screen the radio signals at the frequencies below the plasma frequency $\nu_0=9\times10^{-3}n_e^{1/2}$.

The collisions of giant planets will also generate the bursts of radio waves due to the destruction of their powerful magnetospheres and the magnetic lines reconnections. Let us consider the initial stage of the collision, before the formation of the screening gas clouds. The magnetic field at the visible surface of the Jupiter is $B_i\sim10$~G. The magnetosphere extends for a distance $\sim10^2$ of the Jupiter radius. Note, that at the age of a few Myr the magnetic field of the giant planets was about the order of magnitude larger, in comparison with the old (several Gyr) planets \cite{ReiChr10}. For this reason we may provide the conservative estimation below. The collision of planets excites the plasma turbulence and provides the strong compression of the matter. For the case of the one-dimensional compression, the magnetic field can increase up to the value $B_f\sim10^2(\varkappa/10^2)B_i$. The corresponding energy of this magnetic field
\begin{equation}
E_m\sim\frac{4\pi R_J^3}{3}\frac{B_f^2}{8\pi}\sim6\times10^{34}\left(\frac{\varkappa}{10^2}\right)^2\mbox{~erg}.
\label{emf}
\end{equation}
This energy is released during the time interval $\Delta t \sim R_J/v_{\rm rel}\sim1.5$~hours. Therefore, the radiated power is $P_m\sim \varepsilon E_m/\Delta t\sim10^{31}\varepsilon(\varkappa/10^2)^2$~erg~s$^{-1}$, where $\varepsilon$ is the efficiency of the magnetic to radio energy transformation through the electron cyclotron radiation. If the burst occurred at the distance $l_m$, then the observed flux is
\begin{equation}
L_m\sim \frac{P_m}{4\pi l_m^2}\sim10^{-15}\varepsilon\left(\frac{\varkappa}{10^2}\right)^2\left(\frac{l_m}{10\mbox{~kpc}}\right)^{-2}\mbox{~erg}\mbox{~s}^{-1}\mbox{~cm}^{-2}.
\end{equation}
We assume that the emission of the energy in the plasma in magnetic field occurs at the cyclotron frequency $\omega_c/2\pi = eB_f/(2\pi m_ec)\sim3$~GHz with the spectral density
\begin{equation}
F_m\sim \frac{L_m2\pi}{\omega_c}\sim3\times10^{-2}\varepsilon\left(\frac{\varkappa}{10^2}\right)^2\left(\frac{l_m}{10\mbox{~kpc}}\right)^{-2}\mbox{~Jy},
\label{fmeq}
\end{equation}
This power is strong enough for registration by the modern radio telescopes, which have a typical sensitivity $\sim10$~$\mu$Jy, if $\varepsilon>3\times10^{-4}$. Note that $\varepsilon$ is a free parameter in the current level of the research.

\begin{figure}[t]
\begin{center}
\includegraphics[angle=0,width=0.49\textwidth]{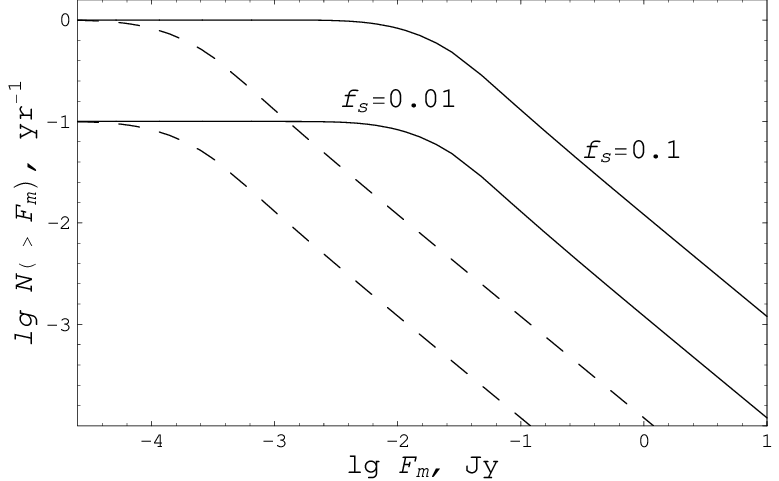}
\end{center}
\caption{The rate of the radio bursts from the collisions of giant planets in the Galactic disk. The logarithm of the number of bursts per year is shown on the vertical axis with the observed flux on the Earth $>F_m$, measured in Jansky. Two solid curves correspond to the number of collision per each star $f_s=0.01$ and $f_s=0.1$ during the time $\sim10^{10}$~yr,  and $\varepsilon=1$ in the both variants. The dashed curves show the same for $\varepsilon=0.01$.} \label{gr2}
\end{figure}

Finally we estimate the distribution of radio bursts observed from the Earth, by taking into account the structure of the Galactic disk. We assume that collisions of the planets are uniformly distributed during the time interval $t_g\sim10^{10}$~yr. Additionally, we will use the standard exponential model of the surface density of the stellar disk:
\begin{equation}
\sigma_s(r)=\frac{M_d}{2\pi r_0^2}e^{-r/r_0}
\label{diskmass}
\end{equation}
where $M_d=8\times10^{10}M_\odot$ and $r_0=4.5$~kpc. In this equation the distance of the star from the center of the Galaxy $r(l,\theta)=[l^2+r_\odot^2-2lr_\odot\cos{\theta}]^{1/2}$ is expressed through its distance from the Earth $l$, where $r_\odot=8.5$~kpc is the distance of the Sun from the Galactic center, $\theta$ is the angle between the directions to the star and to the center of the Galaxy. The requested rate of bursts with the signal, greater than $F_m $, is
\begin{equation}
N(>F_m)=\int\limits_{0}^{2\pi}d\theta\int\limits_{0}^{l_m(F_m)}dllf_st_g^{-1}\sigma_s(r(l,\theta))/m_*
\end{equation}
This rate of bursts is shown at Fig.~\ref{gr2} for $\varepsilon\sim0.01$ and $1$, where $l_m(F_m)$ is the inverse value of (\ref{fmeq}), and $f_s=0.01-0.1$ was evaluated in the Section~\ref{ratesec}. The convergence of the curves at the left side of the Fig.~\ref{gr2} corresponds to the fact, that the sensitive telescope sees all the disk for the both value of $\varepsilon$. Finite disk thickness will modify the high energy limits to $N(>F_m)\propto F_m^{-3/2}$, but these events are rare. In result, we see that the radio bursts are available for the registration, if they could be distinguished from the backgrounds or some other transient radio signals.


\section{Conclusion}
\label{conclsec}

The collisions of giant planets in the process of the planetary system formation are fairly frequent, $\sim 0.001-0.1$ collisions per each planetary system. During these collisions the planets either are merged, lose the energy of relative motion, or  are partially destroyed by losing their outer shells. Collided planets will experience free oscillations with decreasing amplitude and the set of characteristic periods from 20 to 200 minutes \cite{VorZha81}.

Pieces of solid rock from the cores of the giant planets could be formed only in the collisions with the large relative velocities. This is possible only under the exceptional circumstances (on the planet orbits very close to the hosting star or after the supernova explosion). The stony meteorites from the cores of the giant planets differ from the ordinary stony (and iron) meteorites in the sense, that in the interior of giant planets the matter has been exposed to the much higher pressures than in the depths of the terrestrial planets. In addition, in the cores of the giant planets the rocky matter was in contact with the dense hydrogen medium, which could also be reflected in the chemical composition of the meteorites.

There is a physically reasonable hypothesis about the metastability of small pieces of the metallic hydrogen during a very long time interval ($\sim$Gyrs), initiated by the theoretical works on the possibility of the zero-pressure metastable state \cite{BroKagHol71,LifKag72}. In particular, in the framework of this hypothesis the collisionally released fragment of the metallic hydrogen may explain the Tunguska meteorite (M.~N.~Tsymbal and V.~S.~Markin model). Respectively, the combustion of small metallic hydrogen dust particles in the Earth atmosphere would be responsible for the strong ball lightening \cite{NMB}. In the same line of reasoning, the observed rapid disappearance of the circumstellar disk around the Sun-like star TYC824126521 \cite{Meletal12} could be explained by the decay of the ejected metallic hydrogen with a metastability time-scale of the order of several years. These speculations, though sounding exiting, have not the reliable support in the theory of metallic hydrogen, which can not predict exactly the metastability time-scale (review of the problem can be found in \cite{MakShil99}).

Energy release from the radio to optical band will accompany the giant planets collision. The anticipated optical and near UV radiation are the results of transformation of the mechanical energy into the heat with the heating over $10^4$~K. During collisions of giant planets, the collapses of their powerful magnetospheres are possible, accompanied by the generation of strong radio bursts. Our estimations show that one can expect about $\sim0.01-1$ bursts per year with a maximum spectral flux $\sim30 $~mJy at the frequencies $\sim3 $~GHz and with the duration of the each burst $\sim1.5$~hours. To improve the observation signatures it is requested to scan presumably the Galactic regions of young stars because collisions of giant planets take place preferably in the recently formed systems. In addition, one can use the automatic synchronized telescope systems to search for the radio-burst signals after the optical ones.

The authors thank Yu.~M.~Bruk and V.~S.~Markin for useful discussions and the anonymous referees for helpful comments. This study was supported by the grants NSh-871.2012.2 and OFN-17 of the Russian Academy of Sciences.

\end{document}